# Bubble Nucleation on Nano- to Micro-size Cavities and Posts: An Experimental Validation of Classical Theory

S. Witharana[1,2,#], B. Phillips[1], S. Strobel[1], H. D. Kim[1,3], T. McKrell[1], J.-B. Chang[1], J. Buongiorno[1 a)], K. K. Berggren[1], L. Chen[2], Y. Ding[2]

[1] Massachusetts Institute of Technology (MIT), 77 Massachusetts Avenue, Cambridge, 02139 MA, USA

[2] University of Leeds, Clarendon Road, Leeds LS2 9JT, UK

[3] Kyung Hee University, Yongin, Gyunggi 446-701, Republic of Korea

## Abstract

Recently-reported data suggest that bubble nucleation on surfaces with nano-sized features (cavities and posts) may occur close to the thermodynamic saturation temperature. However, according to the traditional theory of heterogeneous bubble nucleation, such low nucleation temperatures are possible only for surfaces with *micro-scale* cavities. Motivated by this apparent contradiction, we have used infrared thermometry to measure the nucleation temperature of water on custom-fabricated nano- to micro-scale cavities (from 90 nm to 4.5 µm in diameter) and posts (from 60 nm to 5 µm in diameter), machined on ultra-smooth and clean silicon wafers using electron beam lithography. Our cavity data are in agreement with the predictions of the Young-Laplace equation, thus re-affirming the correctness of the classic view of heterogeneous bubble nucleation, at least for the water-silicon system investigated here. The data also suggest that individual posts of any size have an insignificant effect on bubble nucleation, as expected from theory.

a)  Author to whom correspondence should be addressed: jacopo@mit.edu, +1(617)253-7316

#) Present address: Max Plank Institute for Solar System Research, 37191 Katlenburg-Lindau, Germany

**Introduction**

Nucleate boiling is an effective heat transfer mechanism, because of the high amount of energy (latent heat) required by the liquid-to-vapor transition. However, a minimum superheat above the thermodynamic saturation temperature of a fluid is required for the initiation of bubble nucleation. In practical applications, it is important to know this minimum nucleation temperature. Lately, several researchers have investigated nano-engineered surfaces as a means to reduce the nucleation temperature and enhance boiling heat transfer. For example, Kim and Vermuri[1] found that the onset of nucleate boiling occurs at 30% lower superheat on alumina nano-porous coatings compared to a plain surface. Nam and Ju[2] observed that the onset of nucleate boiling on their nano-smooth surface (which had no microcavities) occurred at only 9°C above saturation. Ujereh et al.[3] attached arrays of carbon nanotubes to silicon and copper substrates, and used them in boiling experiments with FC-72, to find a heat-transfer-coefficient enhancement of up to 450%. Kim et al.[4] used coatings based on nano-protrusions (nanorods) to create a surface that doubled the value of the critical heat flux for water. Sathyamurthi et al.[5] boiled PF5060 liquid on flat horizontal silicon wafers coated with multi-walled carbon nanotubes and observed CHF enhancement of about 60%. Chen et al.[6] examined boiling and CHF of water on Si substrates covered with Si and Cu nanowires. Bubble nucleation on the nanowire-coated surfaces was achieved at about 10°C above saturation and both CHF and the heat transfer coefficient were more than doubled compared to plain Si. Novak et al.[7] used molecular dynamics to show that nanoscale indentations (order of 1 nm) can promote the formation of vapor nuclei, thus reducing the nucleation temperature from its kinetic limit on an atomically smooth surface.

What makes these findings intriguing is that according to the classic theory of boiling[8,9,10], bubble nucleation from nano-scale structures should not occur at temperatures close to the thermodynamic saturation temperature, but hundreds of degrees above it. Briefly, the Young-Laplace equation describes the pressure difference across the surface of a vapor bubble[11]:

$$P_v - P_l = \frac{2\sigma}{r} \tag{1}$$

where $P_v$ is the vapor pressure, $P_l$ is the liquid pressure, $r$ is the bubble radius of curvature and $\sigma$ is the surface tension. If the equation is solved for $P_v$, then thermodynamic tables can be used to find the corresponding equilibrium temperature for the bubble. For example, for a steam bubble nucleating in water at atmospheric pressure ($P_l$=101 kPa) at a cavity with $r$=10 nm, the vapor pressure is $P_v \approx 4.8$ MPa, and from the steam tables, the temperature of nucleation is ~261°C, which is 161°C above the saturation temperature of water at atmospheric pressure (100°C).

In this work, we have experimentally probed the effect of individual isolated nano- to micro-scale cavities and posts on the bubble nucleation temperature and found that in fact nano-scale cavities require very high superheats for bubble nucleation, while microcavities produce bubble nucleation at much lower temperatures, in good agreement with theory, as explained below.

**Experimental**

Individual isolated boiling-nucleation structures were fabricated on a semiconductor grade, silicon prime wafers. A prime wafer is generally used in semiconductor industry and possesses a very clean surface with surface roughness in the sub-nanometer range (Ra~0.5 nm). We found that a substrate wafer with intrinsic doping (5-25 Ω-cm) and double-side polished is the best choice for determining the surface temperature with an infrared (IR) camera. The wafer thickness was 375±25 µm and the wafers had a static contact angle of approximately 7 degrees with water at room temperature. Artificial nucleation spots were fabricated on these wafers by using electron-beam lithography (EBL), which allowed precise control over the dimensions of the structures. The silicon wafer and a typical fabricated cavity and post are shown in Figure 1. It was of paramount importance to keep all contaminants (e.g. dust) away from the surface, because they could function as nucleation sites and give a spuriously low value of the minimum nucleation temperature. Therefore, the cavities/posts were fabricated in a class-100 (operating at approximately class-10) clean room.

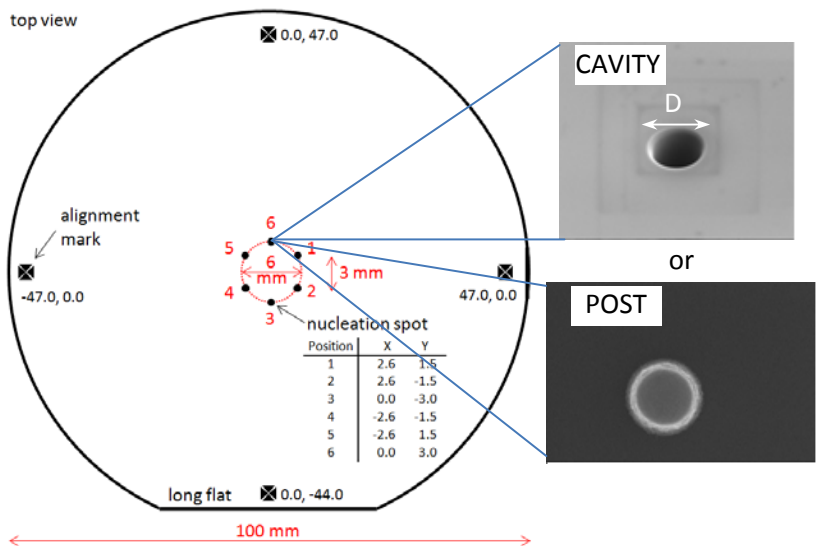

Figure 1: Positions of nucleation spots and alignment marks on a 100-mm-diameter wafer as written by electron-beam lithography. The nucleation spots were arranged on a circle with 6 mm diameter, thus neighboring dots were separated by 3 mm. Alignment marks were placed in all 4 directions at the edge of the wafer. The marks allow orientation and alignment of the wafer for inspection of the nucleation spots after the boiling experiment. A range of post diameters were fabricated and tested across several wafers.

*Fabrication of Posts*

To fabricate raised posts, a clean, out-of-the-box 4" wafer was coated with Hydrogen Silsesquioxane (HSQ), a negative high-resolution electron-beam resist (6% solid, spin-coating at 1 krpm for 60 s resulted in a thickness of about 180 nm). The resist was cross-linked by electron-beam exposure, i.e. the exposed parts of resist were transformed to silicon oxide, which remained on the surface after development. Various diameters of posts were realized by exposing single dots or filled areas. Dots with diameter of 60 nm (single pixel, 230 fC @ aperture 30 µm, and 30 keV acceleration voltage) or circles with larger, variable diameter (2500 µC/cm$^2$ @ aperture 30 µm, pixel raster 20 nm, and 30 keV) were created. The non cross-linked resist was removed by developing in alkaline salt solution (1% NaOH and 4% NaCl in DI water) for 60 s. Subsequently, the wafer was rinsed with deionized water and immediately mounted in the already pre-filled boiling cell (boiling chamber filled with deionized water, semiconductor grade, same as used for flushing the wafer after development). This procedure was followed to avoid contamination and residue on the surface that might be generated in the drying process. The post diameter ranged from 60 nm to 5 µm; the post height was fixed at 180 nm.

*Fabrication of cavities*

For the preparation of cavities, small holes were etched into the surface of the substrate material. PMMA resist was used as etch mask (Polymethyl-Methacrylate A4, spin-coating at 3 krpm for 60 s resulting in a thickness of about 280 nm). In the case of PMMA, a positive electron-beam resist, the EBL exposed areas were removed during development (exposure dose: dot 3 fC, area 400 µC/cm$^2$). The pattern was transferred into the substrate by reactive ion etching (etch gas CF4, 15 sccm, 10 mtorr, and RF-power 110 W resulting in a DC bias voltage of about 270 V). The mouth size of the fabricated cavities ranged from ~90 nm to 4.5 µm. The inner surface of the cavities is covered with fluorinated residue from the fabrication process, which results in a measured static contact angle of 72±4 degrees and advancing contact angle of 76±2 degrees. This relatively high contact angles ensure that vapor can be trapped within the cavities.

*Experimental Apparatus*

Figure 2 shows the experimental setup.  The cylindrical boiling chamber made of polyetherimide had inner dimensions of 25 x 78 mm (height x diameter).  The top of the chamber was sealed with a 3.2 mm thick quartz window, while the bottom surface is the silicon wafer.  Through-holes were provided in the sidewall for vapor escape and liquid refilling.  The wafer was mounted into the chamber in the clean room, again to minimize the possibility of dust deposition on the test surface.  Millipore de-ionized water at atmospheric pressure was the test fluid.  It was degassed by purging Helium gas for 20 minutes while being stirred at 37°C, then sent through a 100 nm filter.  Wafers with cavities and the plain wafers were additionally degassed for approximately 20 minutes by pulling an approximate 25" Hg vacuum on the chamber and heating the chamber to saturation while agitating the water.

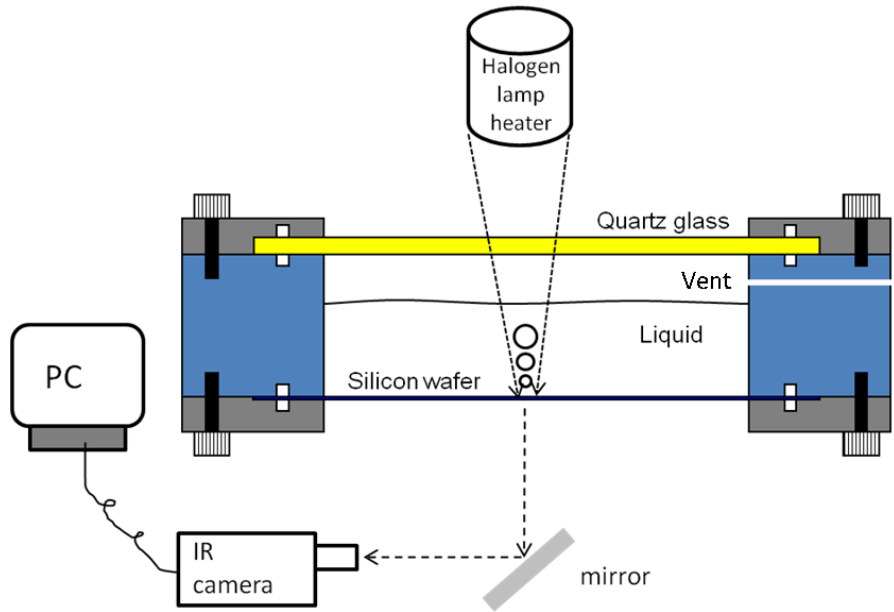

Figure 2:  Schematic diagram of experimental setup for boiling on surfaces with fabricated cavities and posts.  The bubbles are shown to be nucleating near the wafer center, where the lithographic structures were placed.

A Fintech HSH-60/f30 450 W halogen spot lamp mounted above the quartz window was used to heat the wafer surface and induce bubble nucleation non-invasively. The temperature of the wafer surface was measured by an infrared camera (FLIR SC6000, spectral response 3-5 µm) via a gold mirror positioned underneath the silicon wafer. Note that since water is IR opaque while silicon is IR semi-transparent, the camera actually measures the average temperature of a thin (~200 µm) layer of water adjacent to the silicon wafer. IR images of the surface are acquired at 650 to 1000 frames per second and post-processed with ThermoVision ExaminIR, a custom MATLAB script, and ImageJ. Bubble nucleation is evident from the IR data as a sudden drop in the surface temperature, followed by a slow (order of tens of ms) heat up, the so-called waiting time (see Figure 3). The surface temperature right before the drop occurs is assumed to be the nominal temperature of bubble nucleation. The accuracy of temperature reading is ~2°C, as determined through calibration via thermocouples on silicon wafers mounted in a temperature-controlled cell, which was pressurized to prevent boiling. The spatial resolution of the IR camera for this setup is 65 µm.

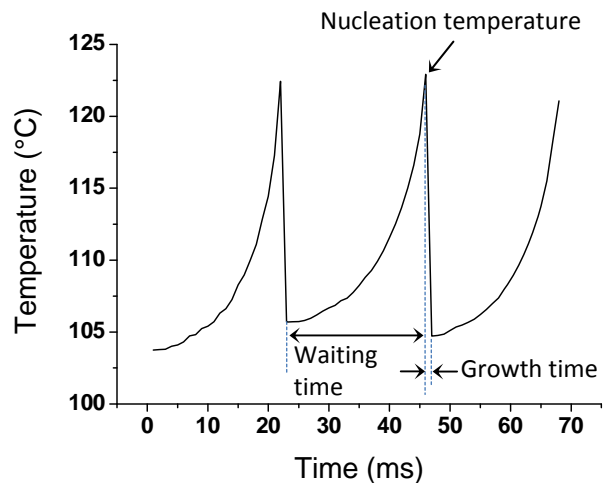

Figure 3: Temperature history for a nucleating cavity of diameter 2.15 µm

**Results**

The test matrix was as follows: 5 smooth wafers (no cavities, no posts), 3 wafers with posts (post diameter: 50, 60, 90, 112, 200, 210, 360, 420 and 5000 nm), and 9 wafers with cavities (cavity diameter: 90, 260, 500, 600, 1450, 1850, 2150, 2350, 4400 and 4550 nm). Each test was repeated a minimum of 3 times. The data are reported in Figure 4, where the bubble nucleation temperature is plotted vs. the cavity diameter. Figure 4 also reports the data for smooth wafers as well as wafers with posts.

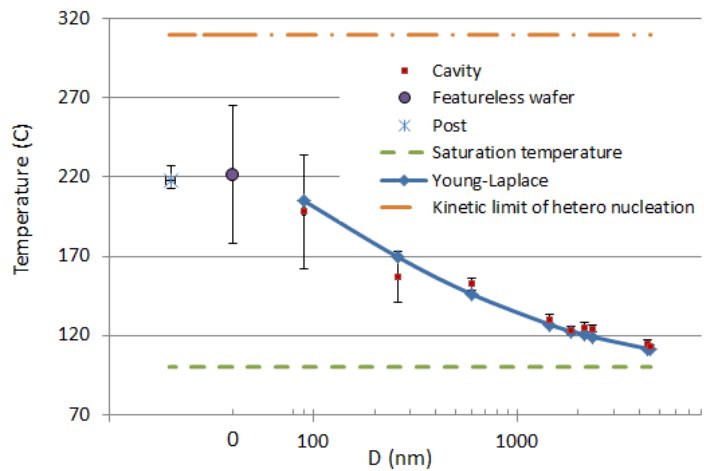

Figure 4: Experimental data for bubble nucleation on surfaces with fabricated cavities and posts. The data shown are mean values of nucleation temperatures measured for multiple bubbles at the same nucleation site, while the error bars indicate the range (min to max). Diameter (D) = 0 indicates data for surfaces with neither cavities nor posts. Data for posts of all sizes are displayed as a single data point.

**Data Analysis**

While there is significant data scattering for the smaller (nanoscale) cavities, the trend is clear: the data over the whole range of cavity sizes explored are in good agreement with the predictions of the Young-Laplace equation (Eq. 1).

Note that the nucleation temperature for the nanoscale cavities approaches the nucleation temperature of wafers with no surface features (neither cavities nor posts), suggesting that the nanoscale cavities are not effective as nucleation sites. This is corroborated by the observation that, in spite of thorough degassing, for the wafers with nanoscale cavities, bubble nucleation occurred randomly on the surface, not preferentially at the locations of the fabricated cavities.

Individual posts of any size (nano- and micro-scale) also seem to have little effect on bubble nucleation, as the nucleation temperature again is close to that of the featureless wafers.  This is expected since, due to their geometry, posts cannot trap the vapor embryos which, according to the classic theory of nucleation, are needed to cause bubble nucleation at lower temperatures.

It is interesting to note that other investigators[2,12] have reported bubble nucleation on nano-smooth heater surfaces to occur at small (5-10°C) superheats.  In light of our findings, we suspect, but cannot prove, that their data might have been "clouded" by the presence of large (micro-scale) contaminants, such as dust, or surface micro-heterogeneities, such as oxide islands[13].  As for the many studies showing low bubble nucleation temperature, high heat transfer coefficient and CHF on nano-engineered surfaces[1,3-6], having established that nano-cavities and nano-posts do not aid bubble nucleation, we must conclude that the low bubble nucleation temperature and high heat transfer coefficient come from the presence of micro-cavities, either pre-existing on the substrate or created (intentionally or fortuitously) by the nano-engineering process (in fact in almost all studies, SEM images of the surfaces show the presence of micro-scale structures and cracks); and the enhanced CHF comes from higher wettability of the nano-engineered surfaces (all studies report low contact angles on their engineered surfaces).

Finally, note that all our data correctly fall between the two theoretical limits, i.e., the thermodynamic saturation temperature of water at atmospheric pressure (100°C), and the kinetic limit for heterogeneous nucleation on a smooth surface. The latter limit was estimated from the equation for rate of vapor embryo formation, $J$ [11]:

$$J = \frac{\rho_{N,l}^{2/3}(1+\cos\theta)}{2F}\left(\frac{3F\sigma}{\pi m}\right)^{1/2}\exp\left(\frac{-16\pi F\sigma^3}{3k_B T_l[\eta P_{sat}(T_l)-P_l]^2}\right)$$

$$F(\theta) = \frac{1}{2} + \frac{3}{4}\cos(\theta) - \frac{1}{4}\cos^3(\theta)$$

$$\eta = \exp\left(\frac{\upsilon_l[P_l - P_{sat}(T_l)]}{RT_l}\right)$$

where $\rho_{N,l}$ is the atom density of the fluid, θ is the static contact angle (7° for water on our silicon wafers), $m$ is the mass of a single molecule of the fluid, $T_l$ is the liquid temperature, $P_{sat}$ is the saturation pressure at $T_l$, $\upsilon_l$ is the liquid specific volume, and $R$ is the gas constant. Following the methodology recommended by Carey[11], $J$ is plotted vs. $T_l$, as shown in Figure 5, from which we estimate the kinetic limit of heterogeneous nucleation to be about 310°C, actually very close to the homogeneous nucleation temperature limit for water.

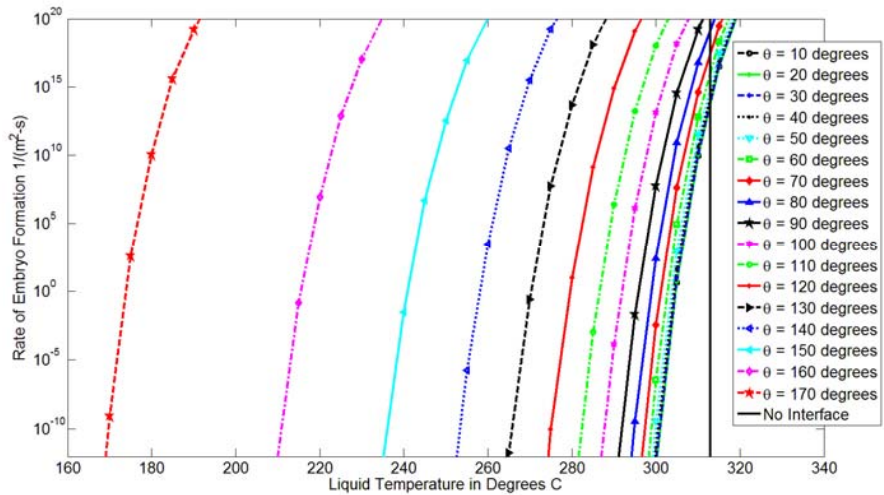

Figure 5: Calculated rate of vapor embryo formation for water on a smooth surface a various contact angles. The "No interface" line represents the homogeneous nucleation limit of water

## Conclusion

Using a combination of a clean test rig and non-invasive data acquisition techniques, we measured the bubble nucleation temperature of cavities and posts with sizes in the nano- to micro-scale range, and found that the traditional description of heterogeneous bubble nucleation, well established in the micro-scale range, holds also at the nano-scale.

## Acknowledgements

Prof. Dimo Kashchiev of the Bulgarian Academy of Sciences for their assistance to SW in fabricating the cavities. The Scanning-Electron-Beam Lithography in the Research Laboratory of Electronics at MIT and the Harvard Centre for Nanoscale Systems for use of their facilities. The MIT Energy Initiative Seed Funds Program for supporting this work. KB and SS also acknowledge support from KACST and Alfaisal University.